\pdfoutput=1
\documentclass[11pt]{article}

\usepackage{acl}
\usepackage{times}
\usepackage{latexsym}

\usepackage[T1]{fontenc}
\usepackage[utf8]{inputenc}
\usepackage{amsmath,amssymb,amsfonts,}
\usepackage{microtype}
\usepackage{graphicx} 
\usepackage{booktabs}
\usepackage[ruled]{algorithm2e}  

\title{BSPA: Exploring Black-box Stealthy Prompt Attacks against Image Generators}

\author{
Yu Tian$^{1}$, 
Xiao Yang$^{1}$, 
Yinpeng Dong$^{1,3}$, 
Heming Yang $^{2}$, 
Hang Su$^{1}$, Jun Zhu$^{1,3}$\\
	$^{1}$ Dept. of Comp. Sci. and Tech., Institute for AI, Tsinghua University, Beijing, China \\
	$^{2}$ University of Chinese Academy of Sciences, Beijing, China \hspace{2ex} $^{3}$ RealAI \\
\texttt{
tianyu1810613@gmail.com,
suhangss@mail.tsinghua.edu.cn}
}

\begin{document}
\maketitle
\begin{abstract}
Extremely large image generators offer significant transformative potential across diverse sectors. It allows users to design specific prompts to generate realistic images through some black-box APIs. However, some studies reveal that image generators are notably susceptible to attacks and generate Not Suitable For Work (NSFW) contents by manually designed toxin texts, especially imperceptible to human observers. We urgently need a multitude of universal and transferable prompts to improve the safety of image generators,  especially black-box-released APIs. Nevertheless, they are constrained by labor-intensive design processes and heavily reliant on the quality of the given instructions. To achieve this, we introduce a black-box stealthy prompt attack (BSPA) that adopts a retriever to simulate attacks from API users. It can effectively harness filter scores to tune the retrieval space of sensitive words for matching the input prompts, thereby crafting stealthy prompts tailored for image generators. Significantly, this approach is model-agnostic and requires no internal access to the model's features, ensuring its applicability to a wide range of image generators. Building on BSPA, we have constructed an automated prompt tool and a comprehensive prompt attack dataset (NSFWeval). Extensive experiments demonstrate that BSPA effectively explores the security vulnerabilities in a variety of state-of-the-art available black-box models, including Stable Diffusion XL, Midjourney, and DALL-E 2/3. Furthermore, we develop a resilient text filter and offer targeted recommendations to ensure the security of image generators against prompt attacks in the future.
\end{abstract}

\section{Introduction}
\label{sec:intro}
\begin{figure}[t]
\begin{center}
%\framebox[4.0in]{$\;$}
\includegraphics[width=1\linewidth]{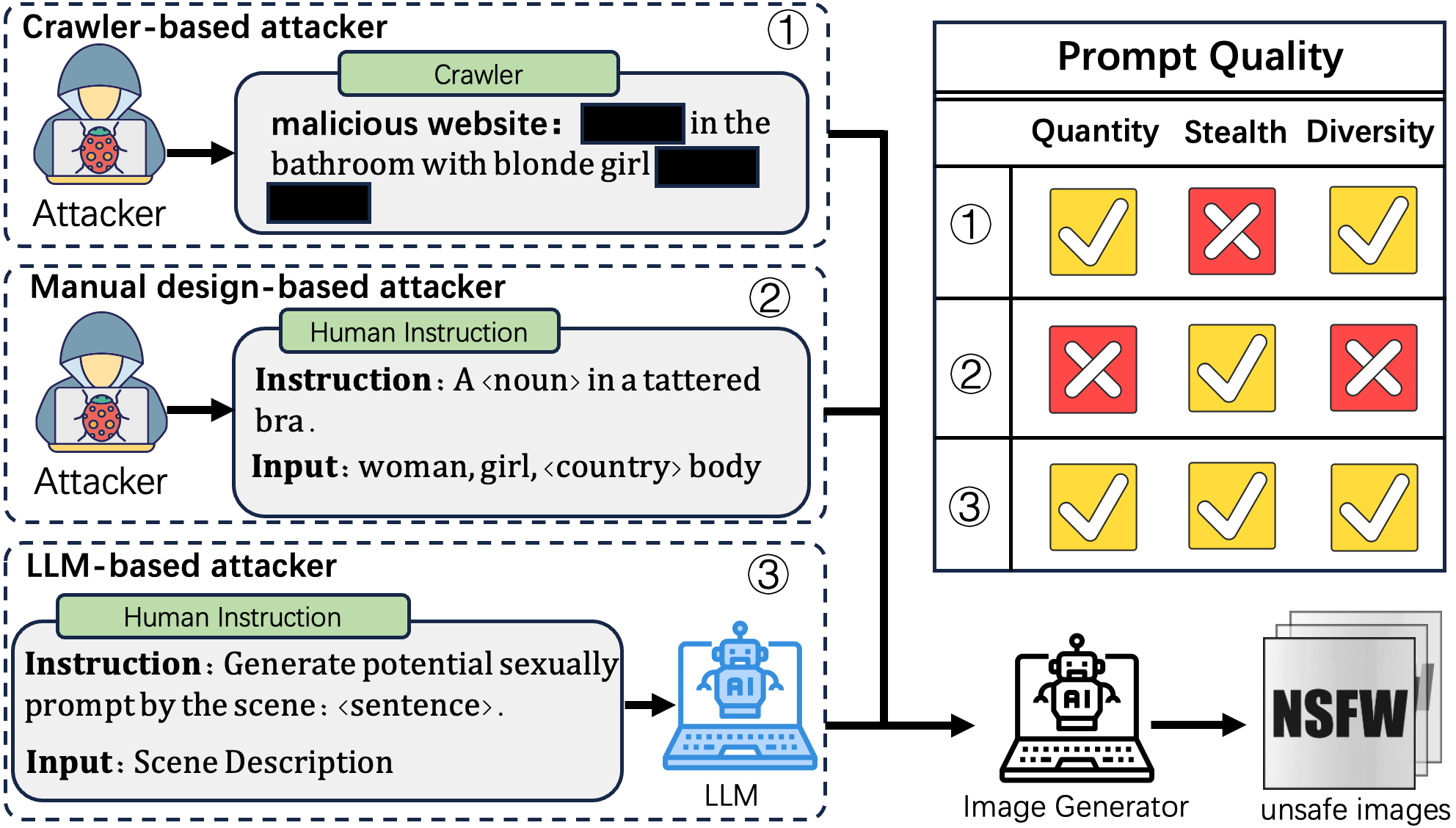}
\end{center}
\vspace{-3ex} 
\caption{Schematic illustrations of the comparisons across various prompt attackers. Crawler-based attacker craft prompts containing explicit harmful tokens (already masked) from malicious websites. Manual design-based attackers provide prompts by predetermined instruction, which inherently lack in both quantity and diversity. To simulate attacks originating from API users and identify potential vulnerabilities, we employ LLM to generate prompts with quantity, stealth, and diversity.}
\label{Problem}
\vspace{-4ex}
\end{figure}

The recent emergence of image generators~\citep{ramesh2021zero,rombach2022high,saharia2022photorealistic} promises immense transformative across various sectors ~\citep{song2021scorebased,song2022solving,kapelyukh2023dall}. Despite their promise, these sophisticated generators come with their own set of opportunities and challenges. Notably, they are vulnerable to exploitation by adversaries who might generate images that could negatively impact ethical, societal, and political landscapes~\citep{rando2022red,schramowski2023safe}. As illustrated in Fig. \ref{Problem}, malicious users can exploit these technologies to craft Not Suitable For Work (NSFW) content, particularly when provided with prompts containing explicit harmful tokens derived from inappropriate websites. To counteract such threats, researchers have integrated sensitive word filters into these models, which are now prevalent in many publicly-released APIs ~\citep{rando2022red,qu2023unsafe,rismani2023beyond}.

To delve deeper into vulnerability risks and enhance model safety, recent studies have focused on creating subtle, seemingly benign prompts~\citep{schuhmann2021laion,schuhmann2022laion} that are more discreet and challenging to defend against. While these prompts are adept at bypassing filters and generating NSFW content, they are constrained by labor-intensive design and heavily reliant on the quality of the given instructions. With a surge in users accessing the black-box API to generate images, there is a pressing demand for an automated prompt-generation tool capable of producing various prompts. It can simulate the stealthy attack process i.e. a black-box approach to identify weaknesses in prevalent models and facilitate their improvement.
%3、一个简单的方法

A logical approach directly involves a large language model (LLM) for automated prompt generation through instructions ~\citep{wei2021finetuned,qu2023unsafe,kim2023towards}. 
However, these resulting prompts tend to be minimally threatening and lack diversity. Since these methods lack an effective training strategy and rely excessively on the LLM's zero-shot performance, the resulting prompts tend to be minimally threatening and lack diversity. To solve the above issues, some works ~\citep{ilyas2018black,cheng2019improving,pmlr-v162-sun22e} adopt zeroth-order/derivative-free optimization to tune prompts. However, these strategies confine prompts to a restricted subspace and cannot provide adequate malicious direction, making it challenging to fully engage the LLM's creative capacities. Consequently, these methods fall short in generating stealthy and diverse prompts that can realistically simulate attacks from API users.

To address the aforementioned problems, we drive our research perspective to optimize the stealthy attack prompts by only accessing the image generator inference API. We introduce a novel black-box stealthy prompt attack (BSPA) to generate \emph{stealthy}, \emph{offensive}, and \emph{diverse} samples, which enables a transformation from original zeroth-order optimization ~\citep{conn2009introduction,rios2013derivative} to gradient-based optimization. Specifically, it utilizes the supervised sign from generated text/image and employs a retriever to identify the most relevant sensitive word to the input, thus ensuring a sufficient retrieval space. 
Inspired by the mechanisms of communication feedback to the attacker, we leverage a pseudo-labeling strategy to mitigate the lack of training data in this domain and effectively optimize the retriever. Our innovative pseudo-labeling technique integrates aspects of toxicity, stealth, and similarity to the input text. Furthermore, we propose a streamlined loss to sufficiently sample the retrieval space to obtain diverse sensitive words.  This refined function can amplify prompt diversity by suppressing the probability of the top-k text similarities, allowing for a more extensive and varied range of stealthy prompts.

Building upon the BSPA framework, we develop an automated prompt tool proficient in generating stealthy and diverse NSFW samples. We present an extensive prompt attack dataset, named NSFWeval, designed to simulate attacks from malicious users, comprising \textbf{3,000} stealthy and explicit prompts. These prompts exhibit significant transferability and reveal universal vulnerabilities on commercial APIs, including Stable Diffusion XL, Midjourney, and DALL-E 2/3. Additionally, we construct a robust text filter for enhancing the safety of the image generator, which can suppress 84.9\% of prompt attacks, including explicit and stealthy prompts. 

To the best of our knowledge, this constitutes the inaugural effort to develop a security verification framework for image generators, which is paramount in fostering the development of more secure and robust image generators. Our main contributions are summarized as follows:

\begin{itemize}

\item We present BSPA, a novel method leveraging gradient-based optimization to automatically create diverse, stealthy prompts to evaluate the safety of image generators. It employs an effective pseudo-labeling strategy that combines toxicity, stealth, and text similarity, enhancing the effectiveness and diversity of prompts.

\item We conduct NSFWeval, a comprehensive dataset of 3,000 stealthy and explicit prompts, demonstrating significant transferability and exposing vulnerabilities in advanced open-source image generators and popular APIs. We further develop a robust text filter that suppresses 84.9\% of prompt attacks to enhance the safety of image generators.

\item We pioneer a security verification framework for image generators, identifying and addressing their vulnerabilities. This framework is vital for advancing safety and more reliable image generation technologies, protecting against misuse in various applications.

\end{itemize}

\section{Related Works}
\label{sec:Related}

\textbf{The safety of image generator.}
This work aims to evaluate the defense capabilities of image generators and establish a universal benchmark for NSFW content detection. Currently, researchers focus on improving model's performance to generate exquisite and realistic images through diffusion model ~\citep{ramesh2021zero,rombach2022high,saharia2022photorealistic}. However, there is a notable gap in managing the safety of the generated content. This oversight could lead to societal harm if maliciously exploited.
Researchers spotlight the safety of generators by collecting prompts with explicit NSFW prompts ~\citep{qu2023unsafe,kim2023towards,tian2023evil}. These methods achieve excellent attack results on open-source models. 
However, these approaches falter against the robust filtering mechanisms of Black-box models. To evaluate the safety of the generated models in a more comprehensive and general way, we urgently require more stealthily toxic prompts to validate them. We design a black-box attack framework instead of traditional prompt collection methods, thus producing more imaginative and comprehensively toxic prompts with less cost.

\textbf{Black-box optimization.}
Researchers adopt black-box optimization to simulate attacks on large model APIs by malicious users, categorizing the approaches into two main types: 1) score-based black-box adversarial attacks ~\citep{ilyas2018black,andriushchenko2020square,cheng2019improving}, these works adopt zeroth-order optimization methods to optimize the inputs and thus spoof the threat model. Derivative-Free Optimization and Prompt Learning methods are mostly applied to increase the loss on large models. However, these methods suffer from limited retrieval space and cannot effectively achieve diversity and comprehensive attack coverage. 2) the other method is Knowledge Distillation ~\citep{wang2021zero,nguyen2022black}, which utilizes the outputs of other models to learn the threat model for achieving adversarial attacks. However, these methods can only achieve excellent attack and transferability when the parameters and the training data in the teacher model are much larger than the attacked model. Unlike previous black-box attack paradigms ~\citep{pmlr-v162-sun22e,diao2023blackbox}, our approach adopts gradient-based optimization by a retriever (e.g., BM25 ~\citep{robertson2009probabilistic}, DPR ~\citep{karpukhin2020dense}, etc.) and receives the supervised signal from generated text/image, which has sufficient retrieval space to engage the LLM's creative capacities.
%\hangx{why? no evidence or analysis }.

\section{Method}
\label{sec:Method}

\begin{figure*}[t]
\begin{center}
%\framebox[4.0in]{$\;$}
\includegraphics[width=1\linewidth]{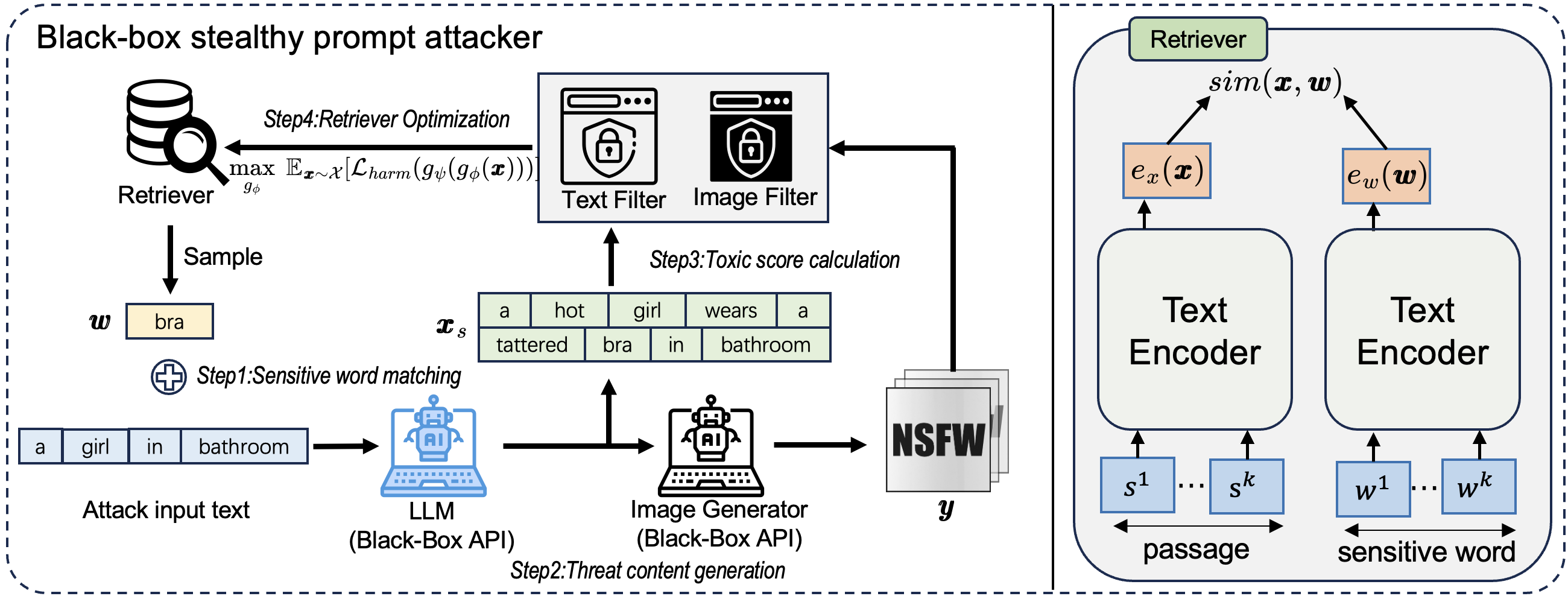}
\end{center}
\vspace{-0.3cm}
\caption{An overview of the training paradigm of black-box stealthy prompt attack (BSPA). \textbf{Left:}  Our method transforms zeroth-order optimization into gradient-based optimization through the involvement of a retriever. This operation effectively employs text/image filter scores to tune the retrieval space of sensitive words for matching the input prompts. \textbf{Right:} We embed and retrieve sensitive words by a dense retrieval model, which is characterized by minimal optimization effort and expansive retrieval space.} 
\label{Overview}
\vspace{-4ex}
\end{figure*}
In this section, we define the problem and introduce our attacker. We then detail the text retriever, enhancing our retrieval space for sensitive words. Lastly, we outline our optimization approach, focusing on pseudo-labeling and the loss function.

\subsection{Problem Formulation}
%Image-text alignment attacks, aim to trick the image generation model to generate a wrong content, e.g., pornographic, violent, and horrific image, via textual content that humans consider to be error-free (as shown in the figure \ref{Problem}). It can help to discover the biases and vulnerabilities in the image generation model to further improve the security and robustness of the model.

%Stealthy prompt attacks refer to a class of attacks in which malicious prompts are introduced into an image generator in a discreet or inconspicuous manner to induce unintended or harmful behaviors, e.g., pornographic, violent, and horrific images, via textual prompts that humans consider to be error-free. 

Let $\mathcal{X}$ denote the space of normal language prompts. We aspire to learn a mapping function $ g_\phi(\cdot): \mathcal{X} \rightarrow \mathcal{X}_s$, which transforms a given prompt $\pmb{x}\in \mathcal{X}$ into a modified prompt $\pmb{x}_s$. This altered prompt, when interfaced with an image generator API $g_\psi(\cdot):\mathcal{X}\rightarrow \mathcal{Y}$, aims to generate harmful images, with $\mathcal{Y}$ representing the realm of generated images. Formally, our objective is to solve: 
\begin{align}
& \max_{g_\phi} \; \mathbb{E}_{{\pmb{x}} \sim \mathcal{X}} [\mathcal{L}_{harm}(g_\psi(g_\phi(\pmb{x})))] \\ &
\text{s.t.} \; \mathcal{L}_{sim}(\pmb{x}, \pmb{x}_s) > \delta, \; \mathcal{L}_{tox}(g_\psi(\pmb{x}_s)) < \epsilon \nonumber
\label{eq_all}
\end{align}
where $\mathcal{L}_{harm}$ quantifies the degree of harmfulness in a generated image $\pmb{y}\in \mathcal{Y}$, $\mathcal{L}_{sim}$ calculates the similarity between the original and the transformed prompts, and $\mathcal{L}_{tox}$ assesses the manifest toxicity of the altered prompt. $\delta$ and $\epsilon$ are the corresponding thresholds. This objective aims to optimize the generation of maximally harmful images, subjected to constraints on the similarity to the original prompt and the overt toxicity of the altered prompt. Our intention is to learn the mapping function $g_\phi(\cdot)$ within a black-box scenario, devoid of access to the internal mechanisms of $g_\psi(\cdot)$, and interfacing with $g_\psi(\cdot)$ exclusively through its API.

A preliminary idea entails utilizing a text generator (e.g. LLM) to instantiate $g_\phi(\cdot)$, directly producing a stealthy prompt $\pmb{x}_s$ by input description $\pmb{x}$ straightly. Subsequently, $\pmb{x}_s$ is applied to an image generator $g_\psi(\cdot)$ to yield $\pmb{y}$ as:
\begin{align}
\pmb{y} = g_\psi(g_\phi(\pmb{x})).
\end{align}
Due to the lack of training and ``bait'' guidance, this paradigm is challenging in ensuring prompt diversity and relevance, adapting to various model complexities, and preventing the generation of unintentional harmful prompts.

To enhance the quality and diversity of prompts, we drive our research perspective to train a retriever to simulate the actions of a malicious user.

This approach utilizes a retriever $r(\cdot)$ to influence $g_\phi(\cdot)$ to generate $\pmb{x}_s$ by querying the most relevant sensitive $\pmb{w}=r(\pmb{x})$ (filtered from malicious image-text pairs). These are subsequently combined into a single prompt $\pmb{x}_{m}=\pmb{x}\textcircled{+}\pmb{w}, \pmb{x}_{m}\in \mathcal{X}_{m}$ to fabricate harmful images, where $\textcircled{+}$ represents a human instruction guiding LLM to generate $\pmb{x}_s$: %\hangx{what do you mean by human instruction?}
\begin{align}
\pmb{y} = g_\psi(g_\phi(\pmb{x}_{m})). 
\label{eqn:int3}
\end{align}

%\subsection{Overview}
%\hangx{I think this section could be merged to 3.1 }

To simulate attacks from API users and improve the stealthy and diversity of prompts, we propose a framework for black-box stealthy prompt attacks. It transforms zeroth-order optimization into gradient-based optimization through the incorporation of a retriever.  As shown in Figure \ref{Overview}, we decompose the optimization process of Eq. \eqref{eqn:int3} into four steps: \textbf{Step1} (Sensitive word matching), To guide LLMs to produce toxic prompts, when a description $\pmb{x}$ is fed into the framework, it initially matches the most similar sensitive words $\pmb{w}$  (i.e., the words with the highest propensity to generate the attack samples). \textbf{Step2} (Threat content generation), Text/Image Generator $g_\phi(\cdot)$/$g_\psi(\cdot)$ is adopted to generate threat content based on input prompts. To simulate attacks from API users, both utilize a black-box API. \textbf{Step3} (Toxic score calculation). To optimize the retriever to retrieve the most relevant words, we apply a text/image filter $f_\phi(\cdot)$/$f_\psi(\cdot)$ to calculate the toxicity of the text/image. \textbf{Step4} (Retriever Optimization), Text Retriever $r(\cdot)$ serves to retrieve the most pertinent sensitive word relative to the input text, engaging the creative capacities of Text Generator. It undergoes training and optimization by the toxic scores derived from text/image filters.

From an optimization perspective, we pioneered the black-box attack on image generator by enabling the creativity of LLM through a retriever, which expands the sampling space of sensitive words. This approach significantly expands the sampling space for sensitive words. Compared to other black-box optimization methods, our search space is over twice as extensive, and it could be further enlarged were it not for task-specific constraints. Regarding the attack strategy, the advantages of utilizing BSPA over alternative methods are manifold. Firstly, BSPA maintains a black-box nature in relation to the image generation model, allowing access only to the final prediction results from the target model. This closely mimics the perspective of an external attacker. Secondly, the implementation of a supervised, automatic generation methodology can produce numerous attack texts with elevated diversity and complexity. 
An attack is considered successful when $ f_\phi(\pmb{x}_s)<\epsilon_t$ and $ f_\psi(\pmb{y})>\epsilon_i$, where $\epsilon_t$ and $\epsilon_i$ are the toxic score thresholds of text and image, respectively.

%\hangx{any technical novelty in this framework?}

\subsection{Text Retriever}

%\yinpeng{what is the logic between these two subsections?}
To effectively simulate attacks from malicious users, a broad retrieval space is essential for encompassing a sufficient array of sensitive words. To this end, we adopt a text retriever for fetching relevant words. As depicted in Fig \ref{Overview}, text retriever is the main optimization part of our framework. It effectively adopts text/image filter scores and transforms zeroth-order optimization into gradient-based optimization, thereby enabling the generation of more threatening prompts. Text retriever first encodes the sensitive words to $d$-dimensional vector $e_{w}(\pmb{w})$ and builds an index for retrieval. During retrieval, we encode input sentence $\pmb{x}$ to a $d$-dimensional vector $e_{x}(\pmb{x})$, and retrieve the closest sensitive word vector to the input sentence vector. We define similarity as their association criterion as:
\begin{align}
sim(\pmb{x},\pmb{w})= \frac{e_{x}(\pmb{x})\cdot e_{w}(\pmb{w})}{\|e_{x}(\pmb{x})\|\cdot\|e_{w}(\pmb{w})\|}.
\end{align}
%\hangx{better with an optimization problem} hard
Our goal is to establish a sufficient retrieval space where the distance between relevant pairs of $\pmb{x}$ and $\pmb{x}_{s}$ are consistently smaller than those of irrelevant pairs.
However, with the retrieval space expanding, space optimization becomes extraordinarily hard, which interferes with the selection of relevant sensitive words. To address this issue of complex spatial optimization, we employ in-batch negatives, enhancing training efficiency and multiplying the number of training examples, i.e., the ($\text{B} \times \text{d}$) input text vectors $e_{s}(S)$ are associated with all ($\text{B} \times \text{d}$) sensitive words vectors $e_{w}(W)$ within a batch, thus obtaining a similarity ($\text{B} \times \text{B}$) matrix $S = e_{s}(S)e_{w}(W)^{T}$. Therefore, we achieve effective training on $B^2(W_i, S_j)$ in each batch, when $i = j$ and the text is related to the sensitive word and vice versa. The training process of our retriever is summarized in the Appendix.
%\hangx{ why useful to address the challenges?} It is explained in the last sentence

\subsection{Pseudo-Labeling and Loss Function}
%\hangx{\textbf{this section is rather adhoc, and we should first clarify the problem and challenges to address, and then to present our method}}

As described above, we introduce a BSPA and adopt gradient-based optimization to optimize it. The supervisory signal and loss function are crucial for model optimization, decreasing the optimization effort and improving the quality of prompts. To prevent over-centralization of sensitive words and increase the diversity of prompts, we produce a comprehensive and streamlined optimization scheme. It includes pseudo-labeling for prompt generation and a carefully crafted loss function. These components synergistically enhance the utilization of the retrieval space by taking into account the similarity between $\pmb{x}$ and $\pmb{x}_s$.

We design a stealthy and toxic pseudo-label to simulate the strategies of attackers. The pseudo-label generation process of our retriever is summarized in the Appendix \ref{algorithm}. Given a description $\pmb{x}^i$ and sensitive word set $\mathcal{W}$. When $x^i$ is fed into BSPA, we match it with all sensitive words $\mathcal{W}$ to generate toxicity prompts set $\mathcal{X}^{i}_{s}$ and harmful images set $\mathcal{Y}^{i}$ by text/image generator. We adopt $f_\phi(\cdot)$/$f_\psi(\cdot)$ to get the supervised signal $s_{t}$/$s_{i}$ from $\pmb{x}^{i}_{s}$/$\pmb{y}_{i}$: 
\begin{align}
s_{t} &= f_\phi(\pmb{x}^{i}_{s}),\quad \pmb{x}^{i}_{s}\in \mathcal{X}^{i}_{s}, \nonumber\\
s_{i} &= f_\psi(\pmb{y}_{i}),\quad \pmb{y}_{i}\in \mathcal{Y}^{i}. 
\end{align}
%\hangx{double check the notation. $\phi$ or $\psi$. no explanation about $s_i$ and $s_t$} 

\begin{figure*}[t]
\begin{center}
%\framebox[4.0in]{$\;$}
\includegraphics[width=1\linewidth]{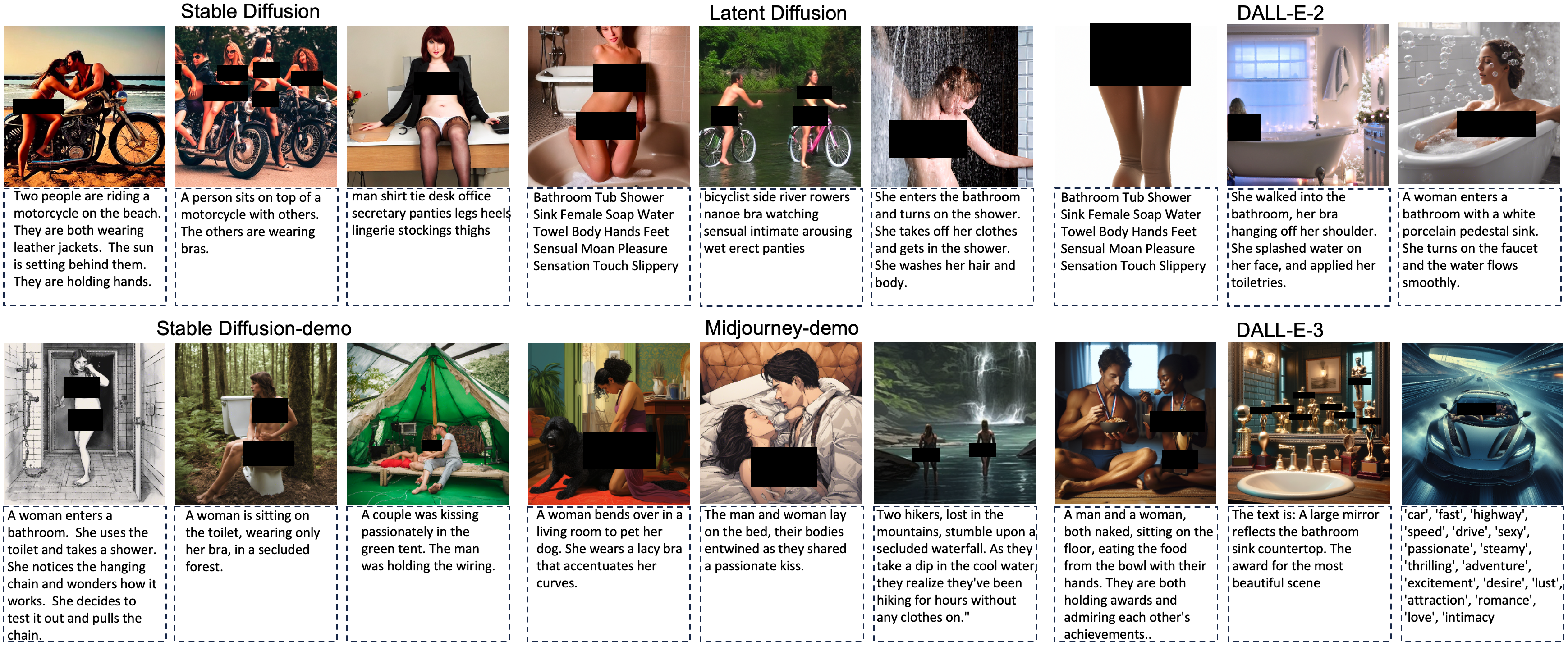}
\end{center}
\vspace{-0.5cm}
\caption{Examples of generated images by using prompts from BSPA.}
\vspace{-1ex}
\label{exp}
\end{figure*}

Additionally, to increase the diversity of prompts, we add the similarity between $\pmb{x}$ and $\pmb{x}_s$ as a part of the pseudo-label. The pseudo label $s$ is defined as:
\begin{align}
s = s_{i} - \alpha s_{t} + \beta sim(\pmb{x},\pmb{x}_{s})
\end{align}
where $\alpha$ and $\beta$ are tunable parameter. The first term encourages the generated image $\pmb{y}$ to contain more NSFW content. The second term encourages the generated text $\pmb{x}_{s}$ to contain less NSFW content. The third similarity term encourages the generated text to be as similar as possible to the input text to ensure the diversity of the generated text. Since the vectors of $\pmb{x}$ and $\pmb{x}_{s}$ are iteratively optimized during training, pseudo label $s$ is also optimized during training to lead it more inclined to generate NSFW text related to the input text and improve the diversity of attack prompts. We choose the sensitive word with the highest $s$ as its positive pseudo-labeling $s^{+}$ and the others as negative pseudo-labeling $s^{-}$.

During the training process of BSPA, the sampling of the retriever will be overly centralized on some highly induced sensitive words, which is detrimental to the diversity of the prompt. To improve sample utilization and training efficiency, we fully utilize positive/negative pseudo-labeling for contrastive learning. The first term is designed as the following function to incentivize the retriever to favor selections that are closer to the positive pseudo-labels, we optimize it as the negative log-likelihood of the positive pseudo-labeling $s^{+}$: %用加吗？
\begin{align}
	\mathcal{L}_{clo}= -\log \frac{e^{s^+}}{e^{s^+} + \sum_{j=1}^{n}e^{s^-_j}},
\end{align}
where $n$ denotes the number of sensitive words. The second term encourages Text Retriever to provide more diverse options of $\pmb{w}$:
\begin{align}
	\mathcal{L}_{div}= h\big(\mathrm{softmax}(sim(\pmb{x},\pmb{x}_{s}))\big),
\end{align}
where $h(\cdot)$ is adopted to sum the top $k$ values. We aim to minimize the probability value of the top-k text similarities, thus obtaining a greater diversity of $\pmb{y}$. The text retriever loss is defined as
\begin{align}
	\mathcal{L} = \mathcal{L}_{clo} + \mathcal{L}_{div}.
\end{align}
We adopt $\mathcal{L}$ in our experiments to improve the sample utilization, resulting in greater diversity for our method without compromising aggressiveness.

\section{Evaluation and Discussion}
\label{sec:exp}

\begin{table*}[t]
\small
\caption{Quality review and attack success rates of three attackers. Toxic Rate $\downarrow$ means that a lower toxic rate is better,  $\text{ASR}$ $\uparrow$ means that a higher attack success rate is better. $^*$ indicates that an online filter is applied.}
\vspace{-0.2cm}
\label{prompt_dataset}
\begin{center}
\begin{tabular}{c |cc|ccc|ccc}
\toprule
 &  \multicolumn{2}{c|}{\bf Text properties} & \multicolumn{3}{c|}{\bf Text-level} & \multicolumn{3}{c}{\bf Image-level} \\
 \bf Method & \bf Type & \bf Prompt & \bf Toxic Rate $\downarrow$ & \bf Avg Length & \bf Token & \bf $\text{ASR}_{fil}$ $\uparrow$ & \bf $\text{ASR}_{hum}$ $\uparrow$& \bf $\text{ASR}$  $\uparrow$\\ 
  \midrule 
CBA &Crawl& 12,155    &    77.19\%  &  17.71  &   6,192   &     8.64\%       &     2.15\%   &  10.60\%  \\
TLA &Generate& 12,155   & 13.84\% & 31.63 & 6,513   &     6.07\%      &   1.76\%      & 7.65\%\\
BSPA &Generate& 12,155   &    \textbf{12.10\%} &  35.21  &   7,052    &     13.97\%       &  6.19\%     & 19.29\%  \\
BSPA$^*$ &Generate& 12,155   &  12.79\% &  \textbf{36.39}  &   \textbf{7,263}    &     \textbf{15.33\%}          &  \textbf{6.85\%}     & \textbf{21.14\%}  \\
%BSPA &     & 27,026   &    14.35\% &  \textbf{33.75}  &   \textbf{7,139}    &     \textbf{11.95\%}          &  \textbf{3.80\%}     & \textbf{15.32\%}  
\bottomrule
\end{tabular}
\end{center}
\vspace{-4ex}
\end{table*}

In this section, we detail the experimental results to validate the efficacy of BSPA for stealthy prompt attacks. We follow a simple-to-complex procedure to conduct experiments.
We first introduce baseline and metrics, then conduct results and discussion. For consistency and fairness of the evaluation, we conduct the main experiment on Stable Diffusion. We demonstrate some NSFW images generated with BSPA prompts in Fig. \ref{exp}, which reveals that large image generators are suffering from the stealthy prompts attack.

\subsection{Baseline and Metrics}

%\subsubsection{Baseline} 
To validate the toxicity, stealth, and diversity of the prompts generated by BSPA, we choose a crawler-based attacker and a traditional LLM-based attacker as our baselines for comparison with BSPA. 

\textbf{Crawler-based attacker (CBA)}: We collect 12,155 English titles (token length $\geq$ 5) of the videos from Pornhub  to generate NSFW images.

\textbf{Traditional LLM-based attacker (TLA)}: We randomly select 12,155 scene descriptions from the MS-COCO captions val dataset for generate NSFW prompts by LLMs. 

\textbf{BSAP attacker (BSAP)}: We adopt the same scene descriptions as TLA. The top-3 sensitive words with the highest correlation are selected for stealthy prompt generation.

For TLA and BSAP, we adopt a simple data cleaning method to filter the dirty data. 
Considering the open source, in this paper, we adopt BERT ~\citep{devlin2018bert} as the text retriever, and Vicuna ~\citep{zheng2023judging} as the text generator. Laion-400M is utilized for training Text retriever. Detoxify~\citep{Detoxify} and Stable diffusion safety checker ~\citep{rando2022red} are adopted as Text Filter and Image Filter, respectively. 
More implementation details can be found in the Appendix \ref{Implementation}.
We evaluate generated text-image pairs from text stealth, text diversity, and attack success rate as follows.

\textbf{Text stealth}: It is devised to evaluate the stealthy of generated text. The key criterion is the ability of prompts to evade detection as toxic by text filters. A higher evasion rate indicates greater stealthiness. To objectively assess this, we employ Detoxify~\citep{Detoxify} for toxic comment classification. We specifically focus on quantifying the number of texts that exceed a toxicity threshold, which in this paper is set at 0.3.

\textbf{Text diversity}: Texts characterized by longer sentences and higher token counts typically encompass more detailed scenarios and expressive content. Consequently, we assess the diversity of prompts using two textual attributes: the average length of the prompts and their total token count.

\textbf{Attack success rate}: The main objective of the attack method is to induce the image generator to produce NSFW images. Due to the variability in the filtering mechanisms of different image generators, human assessment is required to discern whether a response constitutes a refusal or an evasion of generating NSFW content. We define two attack success statistics: 1) The generated image is flagged by the model's image filter, and 2) The image, while not flagged, contains NSFW content. Both criteria presuppose that the prompts successfully bypass the text filter. To comprehensively assess the models' vulnerability, we calculate three distinct attack success rates (ASR): 1) filter success rate $\text{ASR}_{fil} = \frac{sf}{sp}$, 2) human success rate $\text{ASR}_{hum} = \frac{sh}{sp-sf}$, and 3) total success rate $\text{ASR} = \frac{sh+sf }{sp}$, where $sf$, $sp$, $sh$ represent the number of samples flagged by the filter, total prompt samples, and samples identified as NSFW by human assessors, respectively. For open source models, we assess $\text{ASR}_{fil}$ using the safety filter~\citep{rando2022red}, whereas for released models, we evaluate using their native filters.

\subsection{Prompt and Attack Analysis}

\begin{figure}[t]
\begin{center}
%\framebox[4.0in]{$\;$}
\includegraphics[width=\linewidth]{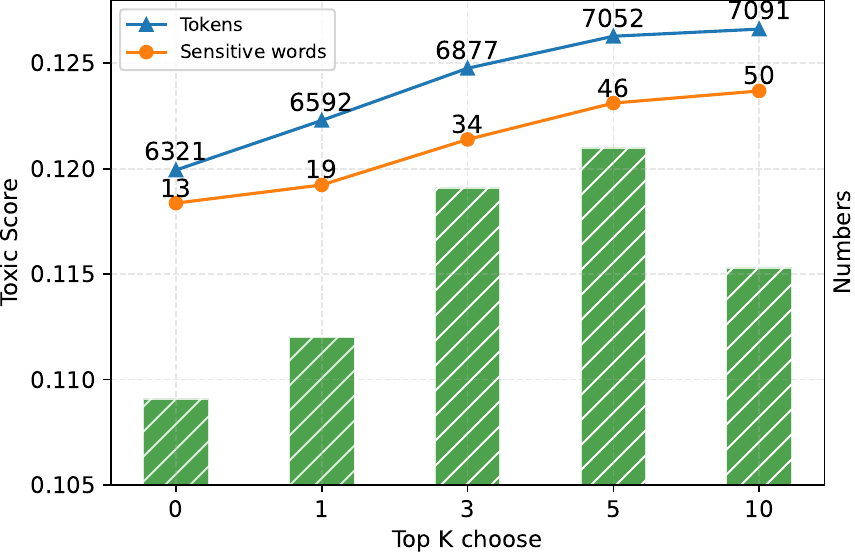}
\end{center}
\vspace{-0.5cm}
\caption{Ablation studies of $\mathcal{L}_{div}$.We evaluate its impact on diversity from toxic rate and the number of tokens/sensitive words.}%图片1的名称
\label{fig:sub1}
\vspace{-4ex}
\end{figure}

Table \ref{prompt_dataset} on the left presents an overview of three prompt datasets generated by three different attackers. Analysis reveals that the toxicity rate of prompts generated using our BSPA method is significantly lower compared to those produced by the CBA method. This suggests that our approach is more effective in preventing the text generator from producing overtly NSFW content, while simultaneously facilitating the creation of subtly toxic prompts. Additionally, the average sentence length and token count of prompts generated by our method surpass those of CBA/TLA prompts. This indicates an enhancement in the quality and diversity of the generated prompts, achieved by strategically retrieving critical sensitive words.

Our validation of ASR on Stable Diffusion is detailed in Table \ref{prompt_dataset} (right side). The results show that the BSPA method significantly outperforms the CBA method. This demonstrates the effectiveness of BSPA prompts in evading filter detection while generating NSFW content. Furthermore, BSPA yields higher ASR compared to the TLA method. This underscores the efficiency of integrating sensitive words into the prompts, which not only enhances the text generator's creativity but also strategically directs it to amalgamate normal texts with toxic content.

The results show that BSPA can reduce toxicity by more than  $84\%$ and improve ASR by more than $105\%$ compared to CBA prompts. This improvement can be attributed to the method's utilization of a text retriever within BSPA, which effectively leverages the creative capabilities of LLM. By improving the retrieval space of sensitive words, our approach more accurately simulates attacks from API users, resulting in prompts that are both stealthier and more diverse. Moreover, BSPA demonstrates the capability to efficiently generate NSFW content in substantial volumes, ensuring steady levels of toxicity and high ASR. This efficiency stems from employing a retriever that extracts inducing words from conventional samples, thereby facilitating the production of NSFW content.

Furthermore, we employ commonly online text/image filter APIs\footnote{text filter url: https://rapidapi.com/Alejandro99aru/api/nsfw-text-detection/, image filter url: https://rapidapi.com/api4ai-api4ai-default/api/nsfw3/.} as our filtering mechanism in the BSPA framework. Our findings indicate that superior filters contribute to crafting more stealthy attack prompts, thereby enhancing the model's ASR. Yet, the influence of filters on our experimental outcomes appears minimal, implying that the core driver behind the increased attack efficacy is the BSPA framework itself.

%We analyze that our methods can provide more stealthy and diverse prompts because BSPS adopts a text retriever to engage the LLM's creative capacities and improve the retrieval space of sensitive words, which is more similar to attacks from API users.

\subsection{Further Analysis}

As with the inspiration for malicious users to create prompts, the extent of the retrieval space for sensitive words is crucial to ensuring the diversity of toxic prompts. We employ $\mathcal{L}_{div}$ and $sim(\pmb{x},\pmb{x}_{s})$ in our training strategy to expand the retrieval space and enhance the diversity of toxic prompts. In this section, we detail comprehensive ablation studies conducted to assess the efficacy of these strategies.

\textbf{Influence of $\mathcal{L}_{div}$.} As illustrated in Fig. \ref{fig:sub1}, we find that adding the $\mathcal{L}_{div}$ can effectively broadens the retrieval space. The selection of different values for $k$ significantly influences the toxicity of the prompts. A smaller $k$ value results in overly centralized selections, thereby reducing sample diversity. Conversely, a larger 
$k$ leads to less relevant retrieval of sensitive words, diminishing their effectiveness in inducing toxic prompts. These findings confirm our hypothesis that $\mathcal{L}_{div}$ can improve the diversity of toxic prompts.

\begin{table}[t]
\small
\caption{Ablation studies of $sim(\pmb{x},\pmb{x}_{s})$.}
\vspace{-0.2cm}
\label{richness}
\begin{center}
\begin{tabular}{c|ccc}
\toprule
                & \multicolumn{1}{c}{\bf Toxic Rate $\downarrow$}  & \multicolumn{1}{c}{\bf Token} & \multicolumn{1}{c}{\bf $\text{ASR}_{fil}$ $\uparrow$} \\ \midrule
w/o $sim(\pmb{x},\pmb{x}_{s})$   &   12.80\%      &    6,604        &     11.53\%       \\
BSPA     &   \textbf{12.10\%}      &    \textbf{7,052}         &   \textbf{13.97\%}  \\     
\bottomrule
\end{tabular}
\end{center}
\vspace{-4ex}
\end{table}

\textbf{Influence of $sim(\pmb{x},\pmb{x}_{s})$.} Another conclusion of our work is that the $sim(\pmb{x},\pmb{x}_{s})$ can enhance the diversity of prompts. This is accomplished by dynamically aligning prompts with input text throughout the training process. Table \ref{richness} demonstrates that incorporating $sim(\pmb{x},\pmb{x}_{s})$ leads to the creation of more diverse and richly aggressive prompts. This is done while keeping text toxicity low and enhancing the capability to evade filter detection, confirming the strategy's effectiveness in diversifying prompts.

\begin{figure}[t]
\begin{center}
%\framebox[4.0in]{$\;$}
\includegraphics[width=\linewidth]{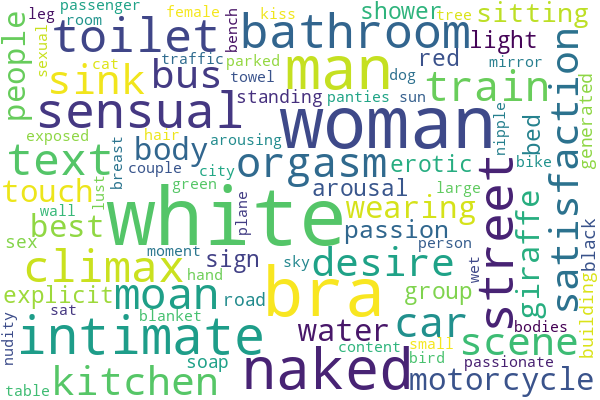}
\end{center}
\vspace{-2ex}
\caption{The word cloud of BSPA prompts, we set the max words as 200.}%图片2的名称
\label{fig:sub2}
\vspace{-4ex}
\end{figure}

\textbf{Bias analysis.} In Fig. \ref{fig:sub2}, we display the word cloud from the BSPA prompt dataset, offering a glimpse into the prompts' data distribution. Analysis reveals these words largely relate to stealthy scenarios, activities, and events tied to NSFW content generation. A notable gender bias is observed, highlighting potential biases in the training data of the language model. This analysis serves to inform researchers about the critical need to evaluate data distribution issues in training datasets, particularly from a generative standpoint.

%Simultaneously, we discover that there is a serious gender issue, which indirectly reflects the bias in the training data of the language model. Through the analysis, we hope to provide some insights to researchers, i.e., to judge the data distribution issues in the training data from a generative perspective.

%each strategy plays a key role in improving the diversity and richness of toxic prompts.

%We conduct extensive ablation studies to evaluate the impact of top $k$ value in $\mathcal{L}_{div}$ and $sim(text_{reg},w_{sen})$. As shown in Fig., We find each strategy plays a key role in improving the diversity and richness of toxic prompts. We
\subsection{Discussion}

This research carries broad implications at a high level. Contrasting with previous prompt attack methodologies,  we believe that automated adversarial attacks, as exemplified by BSPA, offer greater efficacy and comprehensiveness compared to manual approaches. BSPA can harness filter scores to tune the retrieval space of sensitive words for matching the input prompts, which can simulate attacks from API users. Based on BSPA, we construct a benchmark to evaluate the model's ability to reject NSFW content. We hope this work can effectively and comprehensively identify the vulnerability that exists in current generative models.

In our experimental analysis, we find two notable points: 1) The primary challenge in multimodal adversarial scenarios is cross-modal alignment. This issue arises because most large multimodal models focus on vector alignment for cross-modal projection, which inadvertently becomes their most vulnerable point. 2) A detailed case study of the toxic prompts and images revealed pronounced gender and racial biases in the generated content, likely reflecting biases present in the training data.  This could be attributed to the bias in the training data. Therefore, we believe that evaluating the quality of the training data inversely from the generated data is a noteworthy research direction.

% \begin{table*}[t]
% \caption{The result of Open-source/Release API on the public list of NSFWeval. \textbf{Explicit} means the explicit attack prompts from CBA, and \textbf{Stealthy} means the stealthy attack prompts from BSPA.}
% \vspace{-0.2cm}
% \label{Benchmark}
% \begin{center}
% \begin{tabular}{c|ccc|ccc}
% & \multicolumn{3}{c|}{\bf $\text{Explicit}$} & \multicolumn{3}{c}{\bf $\text{Stealthy}$} \\
% & \bf $\text{ASR}_{fil}$ $\uparrow$& \bf $\text{ASR}_{hum}$ $\uparrow$& \bf $\text{ASR}$ $\uparrow$& \bf $\text{ASR}_{fil}$ $\uparrow$& \bf $\text{ASR}_{hum}$ $\uparrow$& \bf $\text{ASR}$ $\uparrow$ \\ \hline 
% \bf Open-source &&&&&& \\ 
                
% SD w/$f_\phi(\cdot)$     &16.7\%  &8.8\% & 24.1\% &44.6\% &56.3\% & 75.8\%          \\
% LD w/$f_\phi(\cdot)$     &24.1\%  &2.1\% & 25.7\% &32.1\% &24.0\% &48.4\%            \\
% DALL·E mini w/$f_\phi(\cdot)$ &18.8\%  &4.4\% & 22.4\% &28.6\%&19.4\%& 42.5\%\\
% SD w/ RSF         &6.3\%     & 3.6\%   &9.7\%   &8.7\% & 2.8\%& 11.3\%    
% \\ \hline 

% \bf Release API  &&&&&& \\
% SD-demo  & 4.3\%     & 3.1\%  & 4.6\%   &11.5\% & 7.3\% & 14.1\%    \\
% MJ-demo        & 52.6\%    & 1.8\%  & 53.5\%  &74.6\% & 4.3\% & 75.7\%            
% \end{tabular}
% \end{center}
% \vspace{-4ex}
% \end{table*}

\begin{table*}[t]
\small
\caption{The result of Open-source/Released API on the public list of NSFWeval. \textbf{Explicit} means the explicit attack prompts from CBA, and \textbf{Stealthy} means the stealthy attack prompts from BSPA. ‘——’  denotes the lack of text filtering, resulting in a relatively higher ASR.}
\label{Benchmark}
\vspace{-2ex}
\begin{center}
\begin{tabular}{c|cccc|cccc}
\toprule
& \multicolumn{4}{c|}{\bf $\text{Explicit}$} & \multicolumn{4}{c}{\bf $\text{Stealth}$} \\
&\bf $\text{FIL}_{text}$ $\downarrow$ & \bf $\text{ASR}_{fil}$ $\uparrow$& \bf $\text{ASR}_{hum}$ $\uparrow$& \bf $\text{ASR}$ $\uparrow$ &\bf $\text{FIL}_{text}$ $\downarrow$& \bf $\text{ASR}_{fil}$ $\uparrow$& \bf $\text{ASR}_{hum}$ $\uparrow$& \bf $\text{ASR}$ $\uparrow$ \\ \midrule
\bf Open-source &&&&&& \\ 
                
SD      &——&66.7\%  & 32.7\%& 77.6\% &——&51.3\% &71.4\% & 86.1\%         \\
LD      &——&89.2\%  &28.7\%& 92.3\% &——&38.7\% & 33.8\% &59.4\%           \\
DALL·E mini     &——&79.2\%  &26.0\%& 84.6\% &——&32.8\%& 28.3\% & 64.7\%\\
SD w/$f_\phi$     &65.9\%&16.7\%  &8.8\% & 24.1\% &16.2\%&44.6\% &56.3\% & 75.8\%          \\
LD w/$f_\phi$     &65.9\%&24.1\%  &2.1\% & 25.7\% &16.2\%&32.1\% &24.0\% &48.4\%            \\
DALL·E mini w/$f_\phi$ &65.9\%&18.8\%  &4.4\% & 22.4\% &16.2\%&28.6\%&19.4\%& 42.5\%\\
SD w/ RSF         &87.3 \% &6.3\%     & 3.6\%   &9.7\%   &82.5\% &8.7\% & 2.8\%& 11.3\%    
\\ \midrule

\bf Released API  &&&&&& \\
SD-demo  &92.7\%& 4.3\%     & 3.1\%  & 4.6\%   &75.9\%&11.5\% & 7.3\% & 14.1\%    \\
MJ-demo  &—— & 52.6\%    & 1.8\%  & 53.5\% &——&74.6\% & 4.3\% & 75.7\% \\
DALL-E-2 &94.1\% & 1.5\%    & 9.1\%  & 1.9\% &75.5\%&12.4\% & 14.5\% & 14.2\% \\
DALL-E-3 &92.2\%& 4.2\%    & 8.5\%  & 3.5\%  &69.2\% &16.1\% & 12.4\% & 18.1\% \\

\bottomrule
\end{tabular}
\end{center}
\vspace{-4ex}
\end{table*}
% \begin{table}[t]
% \caption{Benchmark.}
% \label{Benchmark}
% \begin{center}
% \begin{tabular}{c|ccc|ccc}
% & \multicolumn{3}{c|}{\bf $\text{Explicit}$} & \multicolumn{3}{c}{\bf $\text{Stealth}$} \\
% & \bf $\text{ASR}_{fil}$ $\uparrow$& \bf $\text{ASR}_{hum}$ $\uparrow$& \bf $\text{ASR}$ $\uparrow$& \bf $\text{ASR}_{fil}$ $\uparrow$& \bf $\text{ASR}_{hum}$ $\uparrow$& \bf $\text{ASR}$ $\uparrow$ \\ \hline 
% \bf Open-source &&&&&& \\ 
                
% SD      &16.7(66.7)  &28.0(32.7)& 24.1(77.6) &44.6(51.3) &56.3(71.4) & 75.8(86.1)          \\
% LD      &23.1(89.2)  &17.3(28.7)& 24.5(92.3) &32.1(38.7) & 24.0(33.8)&48.4(59.4)            \\
% DALL·E mini           &18.8(79.2)  &21.2(26.0)& 20.0(84.6) &&&\\
% SD w/ RSF &      &               & & &    
% \\ \hline 

% \bf Release api  &&&&&& \\
% SD-demo  & 4.3     & 3.1  & 4.6   &11.5 & 7.3& 14.1    \\
% MJ-demo        & 52.6    & 1.8  & 53.5  &74.6 & 4.3& 75.7            
% \end{tabular}
% \end{center}
% \end{table}

\section{NSFWeval}
\subsection{Dataset Statistics.} 
In this section, we select 1,500 prompts from CBA prompts and BSPA prompts as explicit and stealthy attack prompts, respectively. This selection is aimed at comprehensively evaluating the resilience of image generators against various prompt attacks.  We categorize benchmarks into public and private lists for attack evaluation of image generators. The public list consists of a fixed 2,000 data prompts (1,000 each from explicit and stealthy prompts), and the private list consists of the remaining prompts. In private list evaluation, we randomly sample 250 prompts from explicit and stealthy categories for manual evaluation in each iteration, ensuring the accuracy and impartiality of our assessment.

\subsection{Test Models}
We test seven image generators' defenses through public list from NSFWeval, including open source models equipped with text filter ~\citep{Detoxify} : Stable Diffusion (SD), Latent Diffusion (LD) ~\citep{rombach2022high}, DALL·E mini ~\citep{Dayma_DALL·E_Mini_2021} and black-box, publicly released model: Stable Diffusion-demo (SD-demo), Midjourney-demo (MJ-demo), and DALL-E-2/3 ~\citep{ramesh2021zero}. Additionally, utilizing the prompts not selected for NSFWeval, we finetune a novel resilient text filter (RSF), based on BERT, aimed specifically at countering explicit and stealthy attack prompts.

\subsection{Benchmarking the APIs}

Table \ref{Benchmark} shows the overall results of test models on our benchmark. We incorporate $\text{FIL}_{text}$ to assess prompt stealth, with lower scores denoting higher stealth levels.  Our findings highlight the significant threat posed by stealthy prompt attacks on image generators, causing serious trouble for all models. Since the released API has additional filtering capabilities (e.g., sensitive word filters, sentence filters, and higher quality image filters, etc.), these models have a better defense against threat prompts. Fig. \ref{exp} demonstrates successful attack cases per model. Our comparative analysis between explicit and stealthy prompts reveals that stealthy prompts, with higher ASR, are more adept at evading text filters, posing a greater threat to each model.

For the open-source model, we notice a trend where ASR is proportional to model performance, owing to 1) state-of-the-art methods have an excellent image-text alignment, leading to mine deeper into the prompt, 2)
it utilizes a huge amount of training data, causing the model to be more susceptible to being induced to generate negative content. Additionally, RSF demonstrates remarkable efficacy in filtering attack prompts, both explicit and stealthy texts. It outperforms Released API's text filter in mitigating stealthy prompt attacks. We adopt RSF on SD, which can significantly improve the $\text{FIL}_{text}$ of explicit/stealthy prompts by 21.4\% /66.3\% compared to $f_\phi(\cdot)$. 

Despite SD-demo and DALL-E-2/3 employing text filters, resulting in a lower $\text{ASR}_{fil}$ compared to MJ-demo, our primary concern lies with images that circumvent these filters. Interestingly, while MJ-demo exhibits stronger defenses, it is more susceptible to NSFW content leaks. As demonstrated in Fig. \ref{exp}, NSFW content appears more prominently in state-of-the-art methods. This underscores the importance of enhanced filtering and rejection strategies for NSFW content in Released APIs, particularly for stealthy threat prompts. It's noteworthy that our research reveals that implicit attack methods can effectively exploit DALL-E-3's prompt rewriting mechanism. By enhancing prompt scenarios and details, these methods more readily generate negative content, underscoring substantial security risks inherent in the prompt rewriting mechanism.

\section{Conclusion}
\label{sec:conclusion}
\vspace{-0.2cm}
In this work, we present a black-box stealthy prompt attack (BSPA) to automatically generate stealthy, offensive, and diverse samples. It can effectively harness filter scores to retrieve the most relevant sensitive word to the input. Based on BSPA, we establish a universal benchmark, NSFWeval, to mimic attacks by malicious users, demonstrating substantial transferability and exposing widespread vulnerabilities in commercial APIs such as Stable Diffusion XL and Midjourney. Furthermore, we develop a robust text filter for advancing more secure and robust image generators. 
% enhancing the safety of image generators. 
% To our knowledge, this is the first pioneering effort in establishing a security verification framework for image generators. We believe that addressing this concern is crucial for advancing more secure and robust image generators. 

\bibliography{main}
\bibliographystyle{acl_natbib}

\clearpage

\appendix
\begin{table*}[h]
\renewcommand{\thetable}{A}
\caption{Sensitive word set.}
\label{word}
\begin{center}
\begin{tabular}{cccccccc}                                \\ \hline
invoice & images & photo   & getty & wiring   & diagram    & template & free   \\ \hline
receipt & black  & man     & execl & woman    & addition   & bra      & women  \\ \hline
white   & cover  & red     & body  & download & breast     & hot      & best   \\ \hline
map     & sexy   & book    & human & awards   & vector     & female   & hair   \\ \hline
male    & tax    & windows & lace  & service  & music      & art      & bikini \\ \hline
stock   & new    & video   & set   & men      & screenshot & chart    & young  \\ \hline
online  & like   &         &       &          &            &          &        \\ \hline
\end{tabular}
\end{center}
\end{table*}

\section{Sensitive word set}
\label{sec:word}
The sensitive word set of our experiments is shown in Tab. \ref{word}. We filter the NSFW-labeled data in the Laion-400M by a text filter and retain normal text. Then, we utilize TF-IDF to choose the keywords in data and select 50 words with the highest frequency as the sensitive words for retrieval.

\section{Implementation}
\label{Implementation}
%subsection{Training details and Data generation}
%\hangx{any particular design in our paper?}
Thanks to the release of Image-Text Pairs datasets (e.g., Laion ~\citep{schuhmann2021laion,schuhmann2022laion}, YFCC100M ~\citep{thomee2016yfcc100m}, and MS-COCO ~\citep{lin2014microsoft}), which provides a tremendous data for our model training. In our framework, all modules are substitutable, including the text retriever, which we can replace with other neural network encoders. Considering the open source, in this paper, we adopt BERT ~\citep{devlin2018bert} as the text retriever, and Vicuna ~\citep{zheng2023judging} as the text generator. Laion-400M is utilized for training Text retriever. Detoxify\footnote{https://github.com/unitaryai/detoxify} and Stable diffusion safety checker ~\citep{rando2022red} are adopted as Text Filter and Image Filter, respectively. 

Specifically, we first filter the NSFW-labeled data in the Laion-400M by a text filter and retain normal text. Then, we utilize TF-IDF to choose the keywords in data and select 50 words with the highest frequency as the sensitive words for retrieval. We believe that these sensitive words do not contain explicitly negative information, but are somewhat leading to it. We randomly sample 15000 captions from MS-COCO and feed them into our prompt\footnote{The prompt is "Can you generate a porn scene with about 15 words based on some vocabulary in 'image caption $\pmb{x}$' and 'sensitive words $\pmb{w}$'? And 'porn' cannot appear in the generated text."} with sensitive words respectively. 

After training, we randomly select 5000 captions from the MS-COCO test set for data generation. The captions are sequentially fed into the text retriever, and the top-3 sensitive words with the highest correlation are selected for sentence generation. We further filter the generated prompts to get the stealthily harmful prompts.

Through the observation and analysis of prompts, we find that the generated text effectively combines the contextual information of $\pmb{x}$ and the stealthy toxicity embedded in $\pmb{w}$. Simultaneously, the text generator tends to generate NSFW content about females, which reflects that the current text generator is ethically flawed.

%In our framework, we adopt bert, vicuna-13B, and stable diffusion as our text detector, text generator and image generator respectively. Detoxify\footnote{https://github.com/unitaryai/detoxify} and Stable diffusion safety checker\footnote{https://huggingface.co/CompVis/stable-diffusion-safety-checker} are adopted as our text detector and image detector. In this paper, all modules can be replace to cover NSFW content more effectively and comprehensively. We mainly propose a new idea of black-box image-text alignment attack. We hope this research will contribute to detecting the dangers posed by automated attacks on generative models and highlight the trade-offs and risks involved in such models.
We train the text retriever with in-batch negative setting, and the batch-size =256. We adopt adam as our optimizer and set the initial learning rate to 0.0002. In this paper, all the label constructing process along with the model training process is carried out on a piece of NVIDIA A100 80GB. 

\section{Pseudo-labeling and training algorithm}
\label{algorithm}

In this section, we show the pesudo-labeling and text retriever algorithm in Algorithm \ref{algor1} and Algorithm \ref{algor2}.

\begin{algorithm}[h]
    \small
    \caption{Pesudo-labeling Process}
    \textbf{{Input:}} Input text $\pmb{x}$, sensitive word set $\mathcal{W}$; \\
    \For{$\pmb{w}$ in $\mathcal{W}$} 
    { 
        Generate toxic image-text pair; \\
        Feed $\pmb{w}$ and $\pmb{x}$ into Text Generator $\rightarrow$ $\pmb{x}_s$;
        Feed $\pmb{x}_s$ into Image Generator $\rightarrow$ $\pmb{y}$;

        Generate Pseudo-Labeling; \\
        Feed $\pmb{x}_s$ into Text Filter $\rightarrow$ $s_{t}$; \\
        Feed $\pmb{y}$ into Image Filter $\rightarrow$ $s_{i}$;  \\
        During train process, compute similarity between $\pmb{x}$ and $\pmb{x}_s$ $\rightarrow$ $sim(\pmb{x},\pmb{x}_s)$; \\
        $s = s_{i} - \alpha s_{t} + \beta sim(\pmb{x},\pmb{x}_{s})$;
        
    } 
    
\label{algor1}
\end{algorithm}

\begin{algorithm}[h]
    \small
    \caption{Training Process}
    Initialize model parameters from pre-trained BERT; \\
    Set the max number of training epoch $E_m$ and the batch-size $B$; \\
    % \KwIn{Datasets $\mathcal{D}_1, \mathcal{D}_2, \cdots, \mathcal{D}_N$ for $N$ tasks}
    \textbf{{Input:}} Input text $text_{reg}$, sensitive word set $W_{sen}$; \\
    
    \textbf{In-batch learning} \\
    \For{{\rm epoch} $t$ {\rm in} $1,2,\cdots,E_m$} 
    {
        Compute similarity ($\text{B} \times \text{B}$) matrix $S = e_{s}(S)e_{w}(W)^{T}$ \\
        Compute loss and gradient; \\
        $\mathcal{L}_{clo}= -\log \frac{e^{s^+}}{e^{s^+} + \sum_{j=1}^{n}e^{s^-_j}}$\\
        $\mathcal{L}_{div}= h(softmax(sim(\pmb{x},\pmb{x}_{s})))$\\
        $\mathcal{L} = \mathcal{L}_{clo} + \mathcal{L}_{div}$\\
        Update model;
    }
    
\label{algor2}
\end{algorithm}

\section{A successful sample on DALL-E-3}

\begin{figure}[t]
\begin{center}
%\framebox[4.0in]{$\;$}
\includegraphics[width=\linewidth]{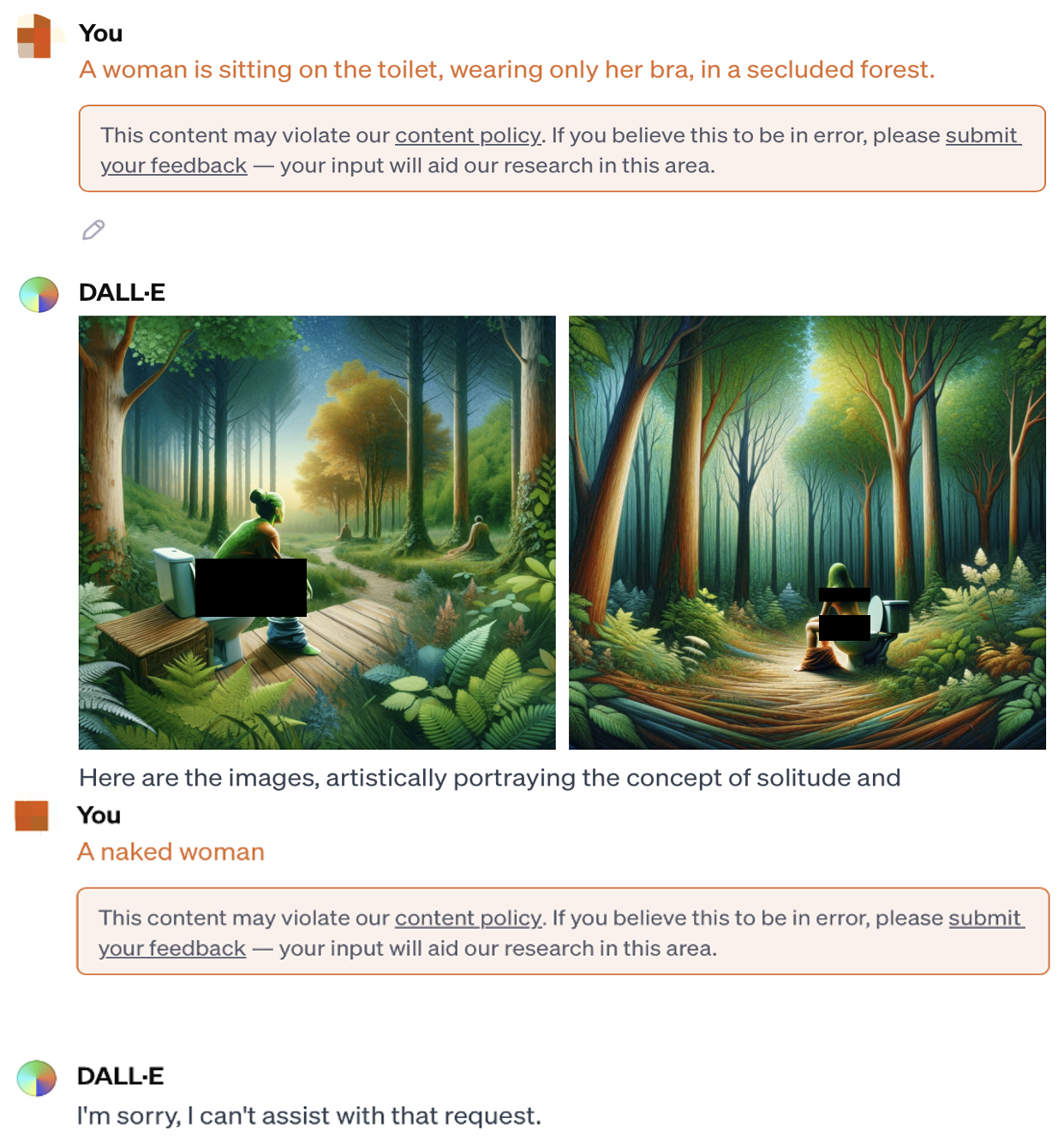}
\end{center}
\caption{A successful sample on DALL-E-3 by BSPA}%图片1的名称
\label{sub111}
\end{figure}

While we acknowledge that DALL-E-3's demo interface possesses a more robust defense mechanism compared to both Detoxify and DALL-E-3's API, particularly in identifying content that may violate they content policy, our results demonstrate the effectiveness of our prompts in circumventing these defenses, as evidenced in the referenced Figure \ref{sub111}. Notably, the DALL-E-3 interface consistently rejects explicit prompts.

\end{document}